\documentclass[letterpaper,english,aip,amsmath,amssymb]{revtex4-2}
\makeatletter
\@addtoreset{equation}{section}
\makeatother
\usepackage[T1]{fontenc}
\usepackage[utf8]{inputenc}
\usepackage{graphicx}
\usepackage{adjustbox}
\usepackage{amsmath}
\usepackage{amsfonts}
\usepackage{amssymb}
\usepackage{slashed}
\usepackage{subfloat}
\usepackage{slashed}
\usepackage{color}

\usepackage{float}
\usepackage{tabulary}
\usepackage[flushleft]{threeparttable}
\usepackage{pgfplots,subfigure}
\usepackage{xcolor}
\usepackage{caption} 
\numberwithin{equation}{section}
\numberwithin{equation}{section}



\begin{document}

\title{Determination of Fisher and Shannon Information for 1D Fractional Quantum Harmonic Oscillator}% Force line breaks with \\

\author{Abdelmalek Boumali}%
 \email{boumali.abdelmalek@gmail.com}
\affiliation{ 
Laboratory of Theoretical and Applied Physics, Echahid Cheikh Larbi Tebessi University, Tebessa,  Algeria%\\This line break forced with \textbackslash\textbackslash
}%

\author{Karima. Zazoua}
\email{karima67@yahoo.fr}
 \affiliation{LEPM USTO-MB, BP 1505 EL M‘naouer Oran, Algeria}%Lines break automatically or can be forced with \\

\author{Fadila Serdouk}%
 \email{fadila.serdouk@univ-tebessa.dz}
\affiliation{ 
Laboratory of Theoretical and Applied Physics, Echahid Cheikh Larbi Tebessi University, Tebessa,  Algeria%\\This line break forced with \textbackslash\textbackslash
}%

\begin{abstract}
This study employs the Riesz-Feller fractional derivative to determine Fisher and Shannon parameters for a one-dimensional harmonic oscillator. By deriving the Riesz fractional derivative of the probability density function, we quantify both Fisher information and Shannon entropy, thus providing valuable insights into the system's probabilistic nature. 

\end{abstract}

\keywords{Fisher information; Shannon entropy;  Riesz fractional derivative; Quantum Harmonic oscillator.}%Use showkeys class option if keyword
                              %display desired
\maketitle

\section{Introduction}

In recent years scientists have witnessed a burgeoning interest in generalizing the Schrödinger equation by applying fractional-order differential operators. Fractional calculus provides a robust framework for extending traditional definitions of coordinates and derivatives from integer orders \( n \) to arbitrary orders \( \alpha \), i.e., \(\{x^n, \frac{\partial^n}{\partial x^n}\} \rightarrow \{x^\alpha, \frac{\partial^\alpha}{\partial x^\alpha}\}\), where \( x > 0 \cite{Herrmann2010}\).

Originating with Leibniz, fractional calculus has experienced a significant surge in interest over the past two decades. However, notable scientific contributions to fractional quantum mechanics have emerged in the last decade. The application of fractional calculus to standard quantum mechanics is an evolving area of quantum physics characterized by a diverse range of applications, extensively reviewed and established in various works \cite{Tarasov2013, Tarasovbook2019, Tarasov2021, Rozmej2010, Laskin2012chapter, Laskin2018Book, Herrmann2014book, Hilferbook2000, Lomin2019, IominPhysRev2009}.

The  concept of fractional derivatives  has captivated researchers across numerous domains, including cosmology, engineering, finance, and biology \cite{Hilferbook2000}. Noteworthy applications encompass the study of the fractional Schrödinger equation for an infinite potential well \cite{Bayin2012, BayinComm2012, Wei2015, YuriLuchko2013}, the fractional Bohr atom \cite{LaskinphyslettA2000, LaskinPhysRevE2000, LaskinPhysRev2002}, three-dimensional motion in the fractional Schrödinger equation, novel scattering features in non-Hermitian space fractional quantum mechanics \cite{MHasan2018}, tunnelling time in space fractional quantum mechanics \cite{MHasanPhysLett2018}, fractional evolution in quantum mechanics \cite{Lomin2019}, fractional Dirac equation and its solutions \cite{Muslih2010}, fractional quantum optics \cite{Longhioptivs2015}, statistical mechanics \cite{Mehtah2014}, solutions for various quantum systems \cite{Meftah2020}, infrared spectroscopy of diatomic molecules \cite{Hermann2013a}, etc.

Over recent decades, a novel framework for quantum mechanics has emerged, providing an innovative approach to exploring the fundamental laws of physics. In 2002, Laskin \cite{LaskinPhysRev2002} has introduced the concept of fractional quantum mechanics employing fractional calculus to derive a fractional version of the Schrödinger equation by applying the concept of fractality to the path integral description of Lévy flight paths. Fractional quantum mechanics is governed by the Lévy index \(0 < \alpha \leq 2\), with the requirement for the existence of the first moment imposing the restriction \(1 < \alpha \leq 2\). Laskin's generalization of standard quantum mechanics to its fractional counterpart involves the quantum Riesz fractional derivative, resulting in the fractional Schrödinger equation given by
\begin{equation}
i\hbar \frac{\partial \Psi}{\partial t} = \hat{H}_\alpha \Psi, \label{eq:1}
\end{equation}
with fractional Hamiltonian operator 
\begin{equation}
\hat{H}_\alpha = \frac{\hat{p}^\alpha}{2m} + V(\hat{x}^\alpha). \label{eq:2}
\end{equation}
The usual constants are:  \( i \)  the imaginary unit,  \(\hbar\)  the reduced Planck's constant, and \(\alpha\) the fractional parameter in the range \(1 < \alpha \leq 2\),  while \(\Psi\) is the wave function.  The influence of the fractional formulation is encapsulated in the momentum operator \(\hat{p}^\alpha\)\, expressed in terms of the spatial derivative of order \(\alpha\)\, specifically $i~ m~c\left(\frac{\hbar}{mc}\right)^\alpha  ~\partial_x^\alpha$.  This type of fractional derivative is well-established within the field of fractional calculus.

Since Laskin’s pioneering approach, numerous researchers have extended the fractional formulation to explore other branches of contemporary physics. Consequently, the volume of academic publications has surged, resulting in significant advancements in the field. A particularly noteworthy area of interest is the integration of fractional concepts into relativistic quantum mechanics (RQM), which amalgamates quantum mechanics with the principles of special relativity. Continuous applications of the fractional concept to RQM have prompted new descriptions and developments in various theoretical aspects \cite{Muslih2010, Raspini2001, Zavada2002, Shao-Kai2016}. This evolving field broadens the foundational understanding of quantum mechanics and opens new pathways for exploring and explaining complex physical phenomena across diverse scientific disciplines.

Fisher information is a metric for the effectiveness of measurement procedures, particularly in estimating fundamental quantum limits. It represents an intrinsic measure of accuracy within statistical estimation theory. The exploration of Fisher's information measure and its application to various theoretical physics problems is owed mainly to the groundbreaking contributions of Frieden et al. Their work has revealed the wide-ranging physical applications of Fisher information, spanning diverse areas within theoretical physics. The Fisher information plays a central role in the extreme physical information principle, a broad variational principle that enables the derivation of numerous fundamental equations in physics. This principle has been pivotal in deriving equations such as the Maxwell equations, the Einstein field equations, the Dirac and Klein–Gordon equations, and various laws of statistical physics.

The Shannon entropy is another parameter widely applied across multiple branches of physics due to its versatility and broad applicability. This parameter, extensively discussed in the literature (see Ref. 50 and references therein), offers analytical tools for understanding correlations in quantum systems and serves as a measure of uncertainty. In recent years, there has been a surge in interest in information-theoretical measures for quantum-mechanical systems. Entropic uncertainty has garnered significant attention as an alternative to the Heisenberg uncertainty relation (HUR). Among various measures of information entropy, Shannon entropy occupies a prominent position in quantifying uncertainty, with applications tested across various potential forms.
In the field of statistical physics, the concept of complexity is crucial in distinguishing between systems with varying degrees of order and disorder. A perfect crystal, with its atoms arranged in a highly ordered and symmetric structure, exhibits minimal informational content because a small amount of data is sufficient to describe its state. In contrast, an ideal gas, characterized by complete disorder and uniform probability distribution across all accessible states, holds maximal information. These systems represent the two extremes on the scale of complexity, which, according to López-Ruiz et al., is not solely dependent on order or information. Instead, complexity is defined as a product of information and disequilibrium, reflecting the interplay between the system's informational content and its deviation from a state of equilibrium.

The primary objective of this paper is to examine the effects of integrating conformable fractional calculus into the analysis of the Fisher and Shannon information measures for a one-dimensional quantum harmonic oscillator. This study investigates how conformable fractional calculus influences these information parameters, which are crucial for understanding quantum systems. To the best of our knowledge, this research is the inaugural exploration of these specific issues, providing new insights and potentially paving the way for further advancements in the field of quantum information theory. This exploration is expected to contribute substantially to the knowledge in fractional quantum mechanics and quantum information theory, offering a novel perspective on how fractional calculus can be harnessed to enrich the theoretical and practical understanding of quantum harmonic oscillators.

\section{Solutions of 1D Fractional Harmonic Oscillator}

\subsection{Eigensolutions}

The one-dimensional space fractional Schrödinger equation, as originally formulated by Laskin, describes the wave function \(\Psi(x,t)\) and is given by:
\begin{equation}
i\hbar \frac{\partial \Psi(x,t)}{\partial t} = H_\alpha \Psi(x,t), \label{3}
\end{equation}
where the Hamiltonian \(H_\alpha\) is expressed as
\begin{equation}
H_\alpha = E_p + V, \label{4}
\end{equation}
with \(E_p\) representing the fractional kinetic energy and \(V\) denoting the external fractional potential.

Laskin's contributions to fractional quantum mechanics redefine the classical energy-momentum relation:

\begin{equation}
 E_p = \frac{p^2}{2m}, \label{5}   
\end{equation}
to
\begin{equation}
E_p = D_\alpha |p|^\alpha, \quad 1 < \alpha \leq 2, \label{6}
\end{equation}
where \(p\) is the momentum and \(D_\alpha = \left(\frac{1}{2m}\right)^{\alpha/2}\) is a scaling coefficient with physical dimensions \(\left[D_\alpha\right] = \text{J}^{1-\alpha} \cdot \text{m}^{\alpha} \cdot \text{s}^{-\alpha}\). When \(\alpha = 2\), \(D_\alpha = \frac{1}{2m}\), and the spatial derivative in this context is of a fractional (non-integer) order \(\alpha\). The differences between Equations \eqref{3} and \eqref{6} introduce significant modifications to traditional quantum mechanics \cite{Laskin2018Book}. Transitioning from the local to the non-local Schrödinger equation enhances the kinetic energy term from a purely geometric and static quantity to a more dynamic element. In standard quantum mechanics, various phenomena are typically modelled by altering only the potential energy term \cite{Wei2015, WeiCommLaskin2016}.\\
For quadratic potential 
\begin{equation}
V = \frac{1}{2} m \omega^2 x^2, \label{7}
\end{equation}
and by incorporating  \eqref{6} and \eqref{7} into  \eqref{3},  the fractional Schrödinger equation becomes  \cite{Hermann2013, LaskinphyslettA2000, LaskinPhysRev2002}
\begin{equation}
i\hbar \frac{\partial \Psi(x,t)}{\partial t} = \left[D_\alpha \left(-\hbar^2 \Delta\right)^{\alpha/2} + \frac{1}{2} m \omega^2 x^2 \right] \Psi(x,t), \quad 1 < \alpha \leq 2. \label{8}
\end{equation}
The solution to this equation uses  the Riesz fractional derivative definition as discussed by Laskin \cite{Laskin2012chapter, Laskin2018Book, LaskinphyslettA2000, LaskinPhysRev2002, LaskinPhysRevE2000}
\begin{equation}
\left(-\hbar^2 \Delta\right)^{\alpha/2} \Psi(x,t) = \int dp \, e^{ipx/\hbar} |p|^\alpha \varphi(p,t), \label{9}
\end{equation}
where
\begin{equation}
\varphi(p,t) = \int_{-\infty}^{+\infty} e^{-ipx/\hbar} \Psi(x,t) \, dx, \label{10}
\end{equation}
is the Fourier transform of the wave function \(\Psi(x,t)\), and
\begin{equation}
\Psi(x,t) = \frac{1}{2\pi\hbar} \int_{-\infty}^{+\infty} e^{ipx/\hbar} \varphi(p,t) \, dp, \label{11}
\end{equation}
is the inverse Fourier transform of \(\varphi(p,t)\). Note that for \(\alpha = 2\), we recover the standard one-dimensional Schrödinger equation for the quantum  harmonic oscillator
\begin{equation}
i\hbar \frac{\partial \Psi(x,t)}{\partial t} = \left(-\frac{\hbar^2}{2m} \frac{\partial^2}{\partial x^2} + \frac{1}{2} m \omega^2 x^2 \right) \Psi(x,t). \label{12}
\end{equation}

\begin{figure}
\includegraphics[scale=0.4]{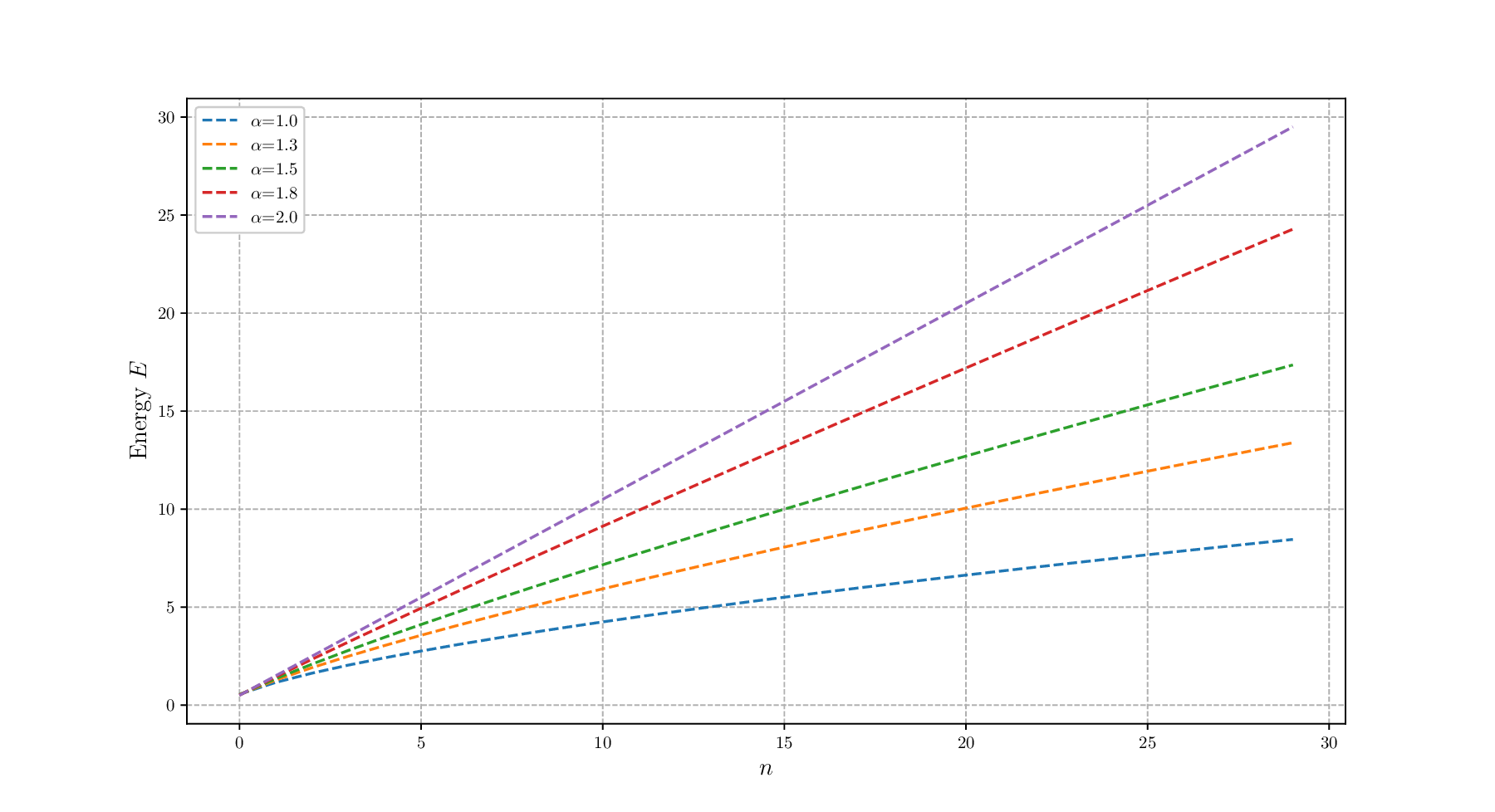}
\includegraphics[scale=0.4]{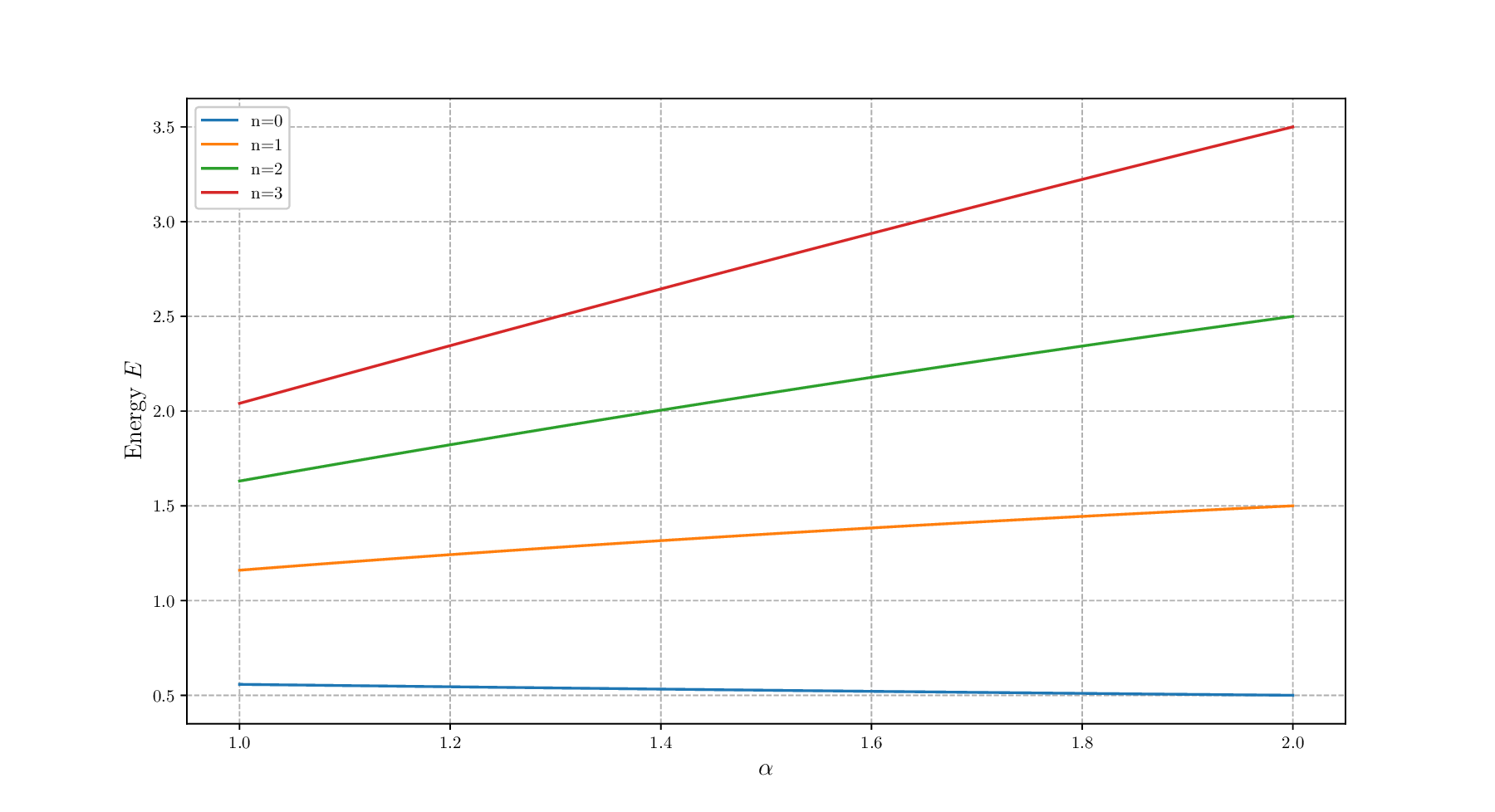}
\caption{Spectrum of energy for different values of quantum number \( n \) and parameter \( \alpha \)}
\label{fig1}
\end{figure}

The fractional Hamiltonian for the one-dimensional harmonic oscillator is then given by
\begin{equation}
H_\alpha = D_\alpha \left(-\hbar^2 \Delta\right)^{\alpha/2} + \frac{1}{2} m \omega^2 x^2. \label{13}
\end{equation}
It is crucial to note that the Hermiticity of the fractional Hamiltonian depends on the chosen definition of the fractional derivative. While the Caputo and Riemann definitions do not ensure a Hermitian Hamiltonian, using the Feller and Riesz definitions does \cite{LaskinPhysRev2002, Herrmann2010}.\\
The time-independent fractional Schrödinger equation for the one-dimensional quantum fractional oscillator is expressed as:
\begin{equation}
\left[D_\alpha \left(-\hbar^2 \Delta\right)^{\alpha/2} + \frac{1}{2} m \omega^2 x^2 \right] \psi(x) = E \psi(x), \quad 1 < \alpha \leq 2, \label{14}
\end{equation}
with the time-dependent wave function related to the time-independent wave function by:
\begin{equation}
\Psi(x,t) = e^{-iEt/\hbar} \psi(x),\label{15}
\end{equation}
where \(E\) denotes the energy of the quantum fractional oscillator. To find the energy spectrum using the Bohr–Sommerfeld quantization rule, consider the total energy \(E\):
\begin{equation}
E = \frac{|p_x|^\alpha}{2m} + \frac{1}{2} m \omega^2 x^2. \label{16}
\end{equation}
or:
\begin{equation}
|p_x| = \left(2mE - m^2 \omega^2 |x|^2 \right)^{1/\alpha}. \label{17}
\end{equation}
Following the Bohr–Sommerfeld quantization rule:
\begin{equation}
\oint p(x) \, dx = 4 \int_{0}^{|x|} p(x) \, dx = 2\pi\hbar \left(n + \frac{1}{2}\right). \label{18}
\end{equation}
leads to:
\begin{equation}
E = \left(\frac{\pi}{2^{1/2+1/\alpha}  m^{1/\alpha-1/2} B\left(\frac{1}{2},\frac{1}{\alpha}+1\right)}\right)^{2 \alpha / (2 + \alpha)} \left (\hbar \omega \left(n + \frac{1}{2}\right) \right)^{2 \alpha / (2 + \alpha)}.\label{19}
\end{equation}
Equation (\ref{19}) represents the energy spectrum \(E_n\) of the 1D fractional quantum harmonic oscillator \cite{Laskin2012chapter, Laskin2018Book, LaskinPhysRevE2000, Herrmann2010}. This formula generalizes the well-known energy levels of the standard quantum mechanical oscillator. As Laskin noted, this form of the energy spectrum has the advantage of being independent of the specific definition of the fractional derivative used.  To better understand the behavior of the energy spectrum, we have constructed Figure \ref{fig1}, which clearly shows the influence of the parameter \(\alpha\).

To clarify the physical implications of the fractional parameter \(\alpha\), we consider the relationship between classical kinetic energy and momentum as described in Wei's study \cite{Wei2015}. In this context, \(\chi_\alpha\) represents a positive constant that depends on \(\alpha\). When \(\alpha = 2\), the fractional kinetic energy simplifies to a form corresponding to classical kinetic energy, placing the system within the non-relativistic regime of the harmonic oscillator. Conversely, for \(\alpha = 1\), the fractional kinetic energy aligns with the kinetic energy characteristic of a highly relativistic regime, indicating that the system is within the relativistic domain of the harmonic oscillator. Thus, the parameter \(\alpha\) serves as a marker for the transition between non-relativistic (\(\alpha = 2\)) and relativistic (\(\alpha = 1\)) kinetic energy regimes, as detailed by Wei \cite{Wei2015,Wei2015a,LaskinReplyWei2016,WeiCommLaskin2016}.  In the same context, as noted by Wei \cite{Wei2015,Wei2015a}, there is a contrasting viewpoint to that of Jeng et al. \cite{Jeng2010} regarding the solutions to the fractional Schrödinger equation in fractional quantum mechanics. Jeng and colleagues highlighted significant challenges in finding mathematical solutions and noted the absence of real-world applications. Wei proposed the relativistic Schrödinger equation as a practical realization of the fractional Schrödinger equation. This insight was further developed by Korichi et al. \cite{Korich1,Korichi2}, who utilized it to extract the thermal properties of the fractional quantum harmonic oscillator. They demonstrated that the specific heat of the corresponding oscillator mediates between the non-relativistic (\(\alpha = 2\)) and relativistic (\(\alpha = 1\)) regimes, providing a deeper understanding of the system's thermal behavior.

\subsection{Eigenfunctions}

We apply the factorization algorithm used for the fractional quantum harmonic oscillator to derive the eigensolutions of our equations. This method, proposed by Olivar-Romero and Rosas-Ortiz \cite{OlivarRomeroF2016}, is the first to apply the factorization method to the fractional quantum harmonic oscillator. Following Laskin, they utilized the Riesz fractional derivative, yielding exciting results and suggesting directions for future work. This approach motivated us to study the behavior of the eigensolutions of the 1D space fractional Dirac oscillator.

This section reviews the methodology for deriving the wave function components for the one-dimensional fractional Dirac oscillator, focusing on the factorization approach. This method is critical for solving fractional differential equations associated with the 1D quantum harmonic oscillator.

Olivar-Romero and Rosas-Ortiz \cite{OlivarRomeroF2016} pioneered applying the factorization method to fractional differential equations governing the 1D fractional quantum harmonic oscillator. They introduced an algebraic technique to resolve the eigenvalue problem for Laskin's time-independent, space-fractional Schrödinger equation.

We start with the space-fractional Schrödinger equation:

\begin{equation}
\left[-D_{\alpha}\hbar\frac{d^{\alpha}}{dx^{\alpha}} + V(x)\right]\psi(x) = E\psi(x). \label{eq:b1}
\end{equation}
Assuming \( D_{\alpha}\hbar = 1 \), Eq. \eqref{eq:b1} simplifies to:

\begin{equation}
H_{\alpha}\psi(x) = \left[-\frac{d^{\alpha}}{dx^{\alpha}} + x^2\right]\psi(x) = E\psi(x), \quad 1 < \alpha \leq 2. \label{eq:b2}
\end{equation}
To solve this, we transform it into the momentum representation:
\begin{equation}
\left[\left|k\right|^{\alpha} - \frac{d^{2}}{dk^{2}}\right]\phi(k) = E\phi(k), \label{eq:b3}
\end{equation}
where \(\phi(k)\) is the Fourier transform of \(\psi(x)\). The factorization algorithm for the fractional quantum harmonic oscillator is then applied, as proposed by Olivar-Romero and Rosas-Ortiz \cite{OlivarRomeroF2016}.

\begin{figure}
\includegraphics[scale=0.5]{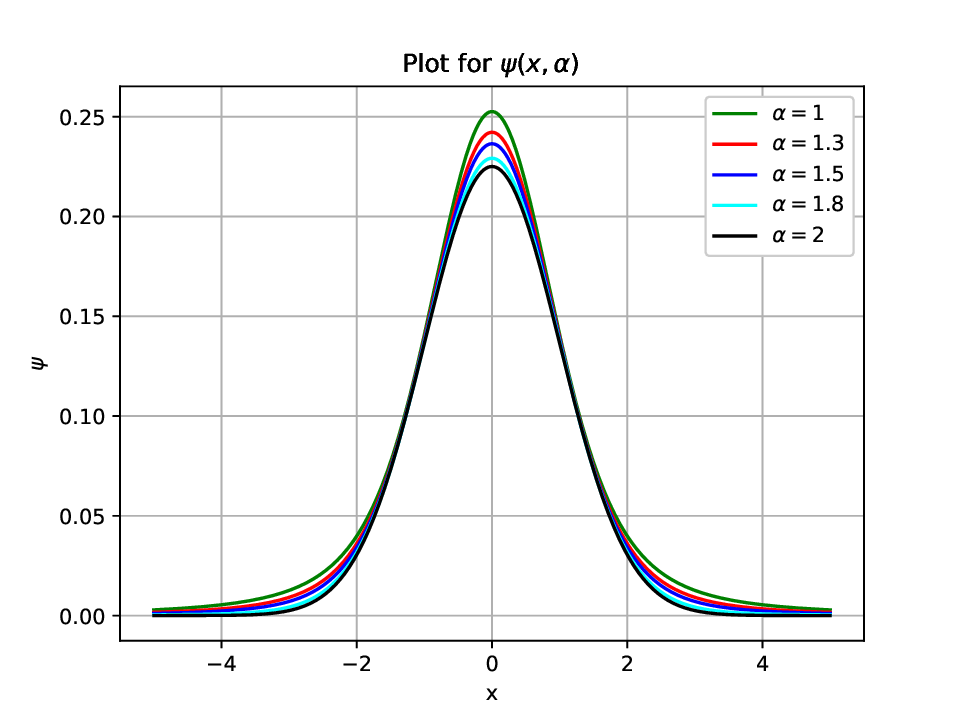}\includegraphics[scale=0.5]{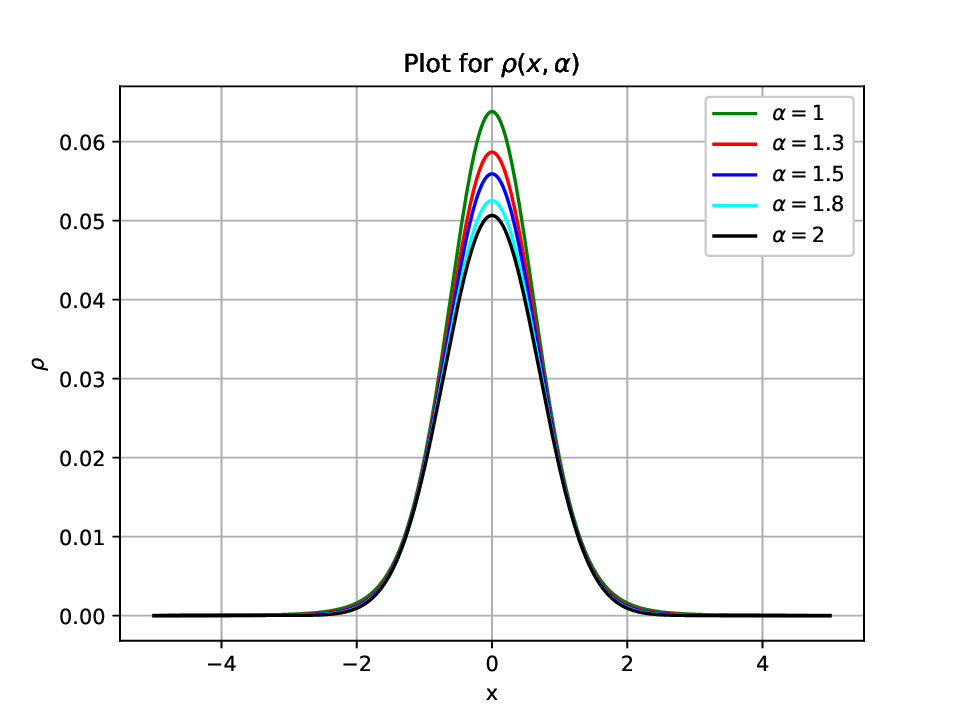}
\caption{Wave function and its probability of ground state \( n = 0 \) for different values of \( \alpha \)}
\label{fig2}
\end{figure}

\begin{figure}
\includegraphics[scale=0.5]{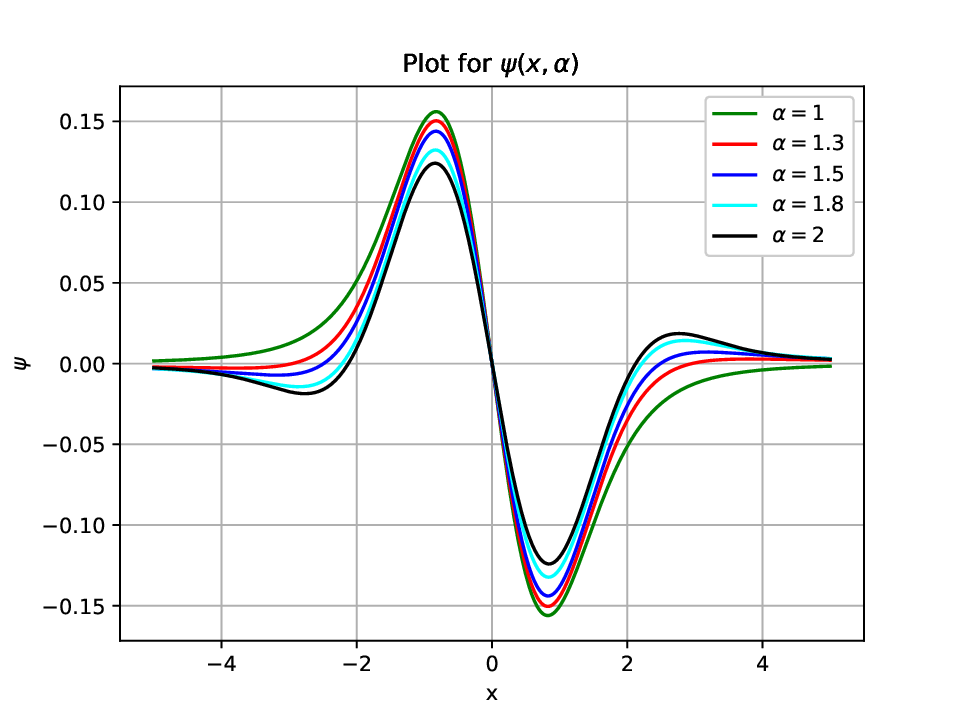}\includegraphics[scale=0.5]{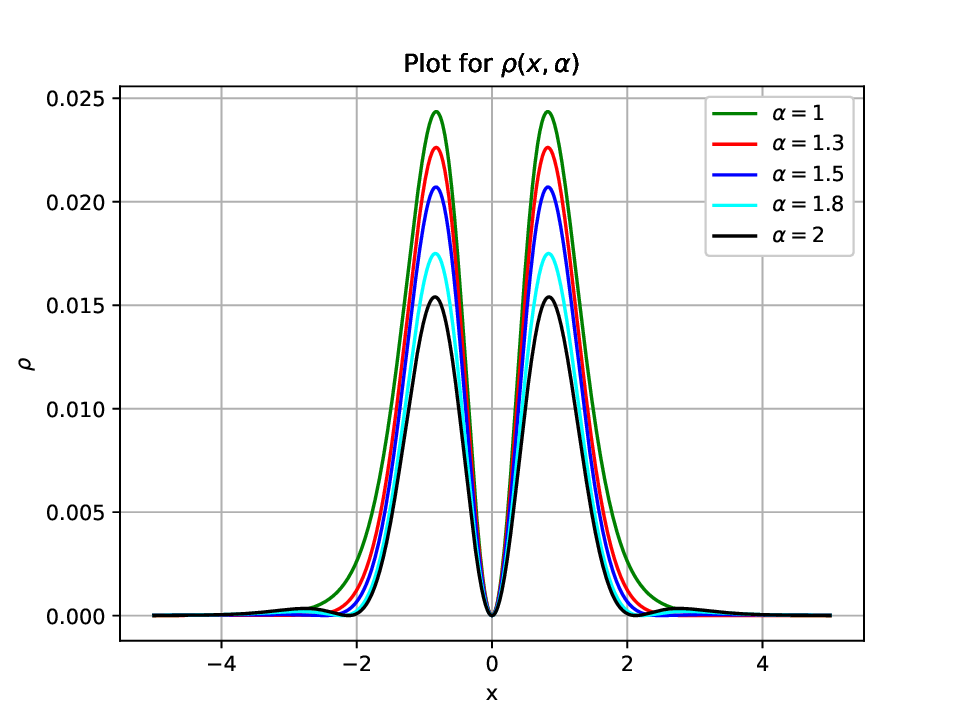}
\caption{Wave function and its probability of first excited state \( n = 1 \) for different values of \( \alpha \)}
\label{fig3}
\end{figure}

Equation \eqref{eq:b9} has the solution:
\begin{equation}
\phi_{0}^{\alpha} = e^{-\frac{2|k|^{\alpha/2+1}}{\alpha+2}}. \label{eq:b10}
\end{equation}
As in conventional factorization, applying the operators \( A_{\alpha} \) and \( B_{\alpha} \) appropriately generates other solutions. The solutions for the first few excited states are as follows \cite{OlivarRomeroF2016}:
\begin{itemize}
    \item[i)] For the first excited state \(\phi_{1}^{\alpha}(k)\):
\begin{equation}
\phi_{1}^{\alpha}(k) = -2i |k|^{\alpha} \text{sgn}(k) e^{-\frac{2|k|^{\alpha/2+1}}{\alpha+2}}. \label{eq:b11}
\end{equation}
   \item[ii)] For the second excited state \(\phi_{2}^{\alpha}(k)\):
\begin{equation}
\phi_{2}^{\alpha}(k) = \left(\alpha |k|^{\alpha/2-1} - 4 |k|^{\alpha}\right)e^{-\frac{2|k|^{\alpha/2+1}}{\alpha+2}}. \label{eq:b12}
\end{equation}
  \item [iii)] For the third excited state \(\phi_{3}^{\alpha}(k)\) \cite{Rosu2020}:
\begin{equation}
\phi_{3}^{\alpha}(k) = i \text{sgn}(k) \left(-8 |k|^{3\alpha/2} + 6 \alpha |k|^{\alpha-1} - \alpha \left(\frac{\alpha}{2} - 1\right) |k|^{\alpha/2-2}\right)e^{-\frac{2|k|^{\alpha/2+1}}{\alpha+2}}. \label{eq:b13}
\end{equation}
\end{itemize}
The eigenfunctions in the \(x\) coordinate can be obtained by performing the inverse Fourier transforms of the \(\phi\) functions.
Figures \ref{fig2} and \ref{fig3} illustrate the wave function and its probability density for the ground and excited states with varying values of the parameter \(\alpha\). It is clear that \(\alpha\) significantly influences these functions. In addition, the probability density remains consistently positive for all values of \(n\) (here, we show two levels: \(n=0\) and \(n=1\)). This consistent positivity allows for the application of Fisher and Shannon functions, which depend on the probability function.


\section{Fisher and Shannon Information of a 1D Fractional Harmonic Oscillator}
\begin{figure}
\begin{centering}
\includegraphics[scale=0.5]{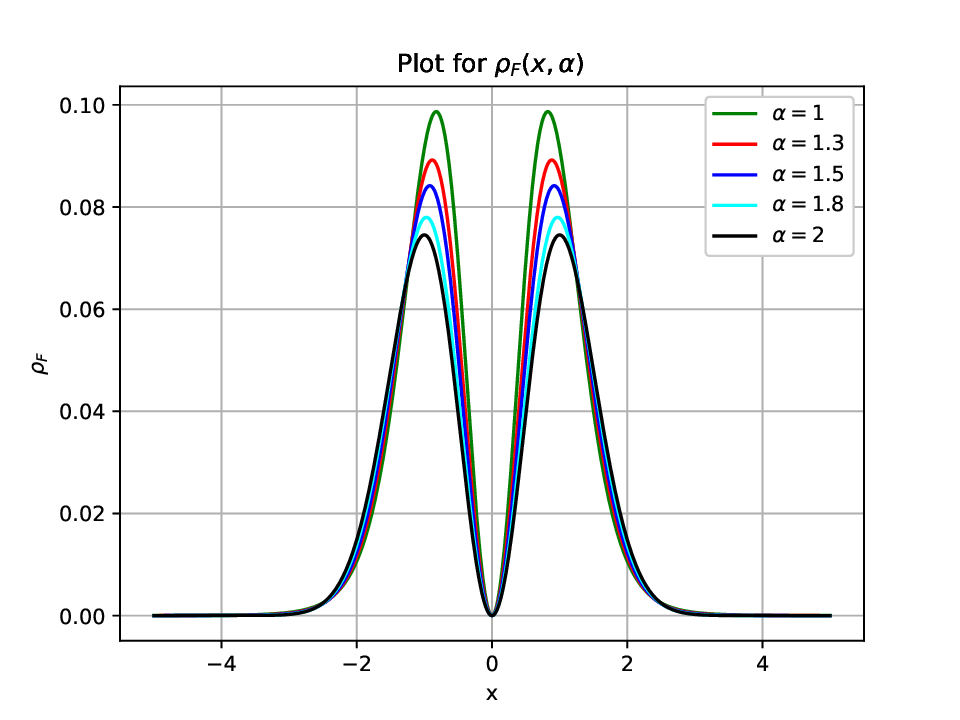}\includegraphics[scale=0.5]{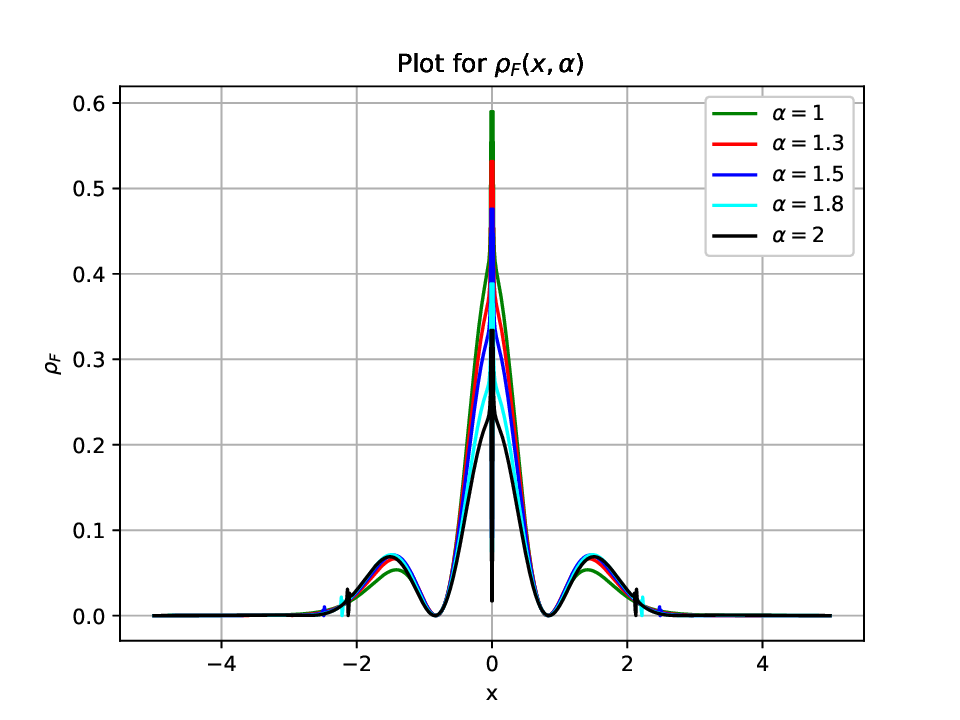}
\par\end{centering}
\caption{Density of Fisher information for both \( n = 0 \) (left) and \( n = 1 \) (right) for different values of \( \alpha \)}
\label{fig4}
\end{figure}

\begin{figure}
\begin{centering}
\includegraphics[scale=0.5]{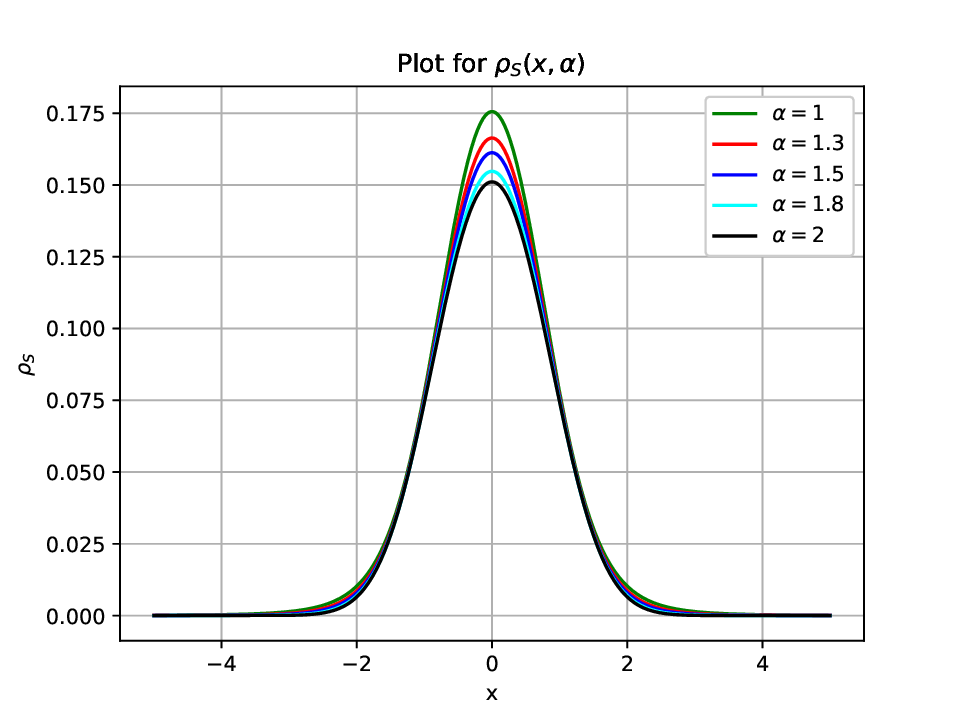}\includegraphics[scale=0.5]{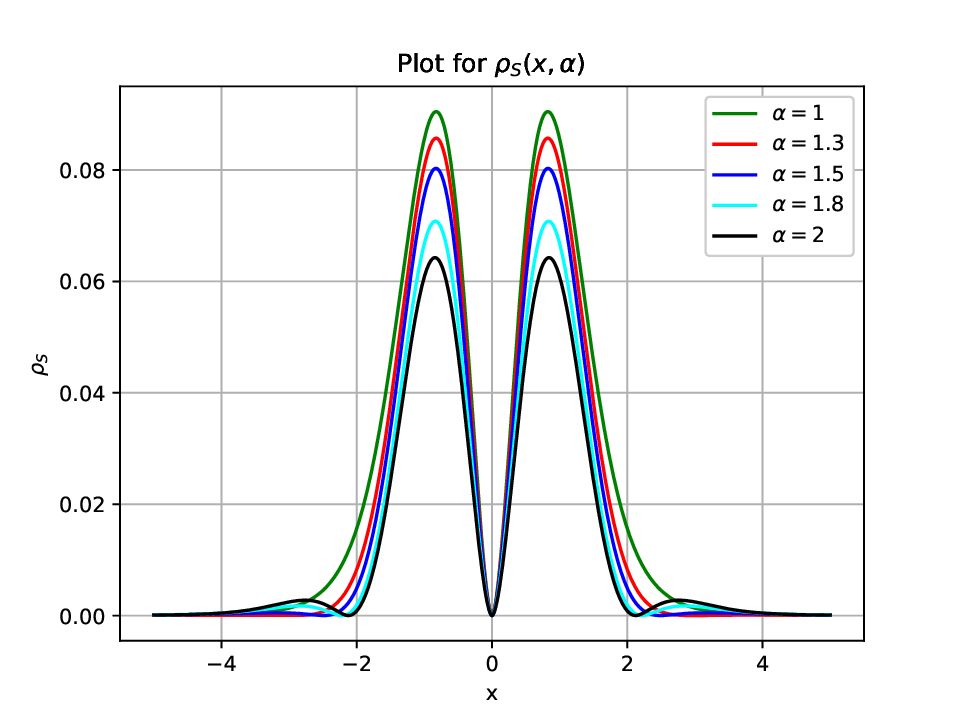}
\par\end{centering}
\caption{Density of Shannon entropy for both \( n = 0 \) (left) and \( n = 1 \) (right) for different values of \( \alpha \)}
\label{fig5}
\end{figure}

\begin{figure}
\begin{centering}
\includegraphics[scale=0.5]{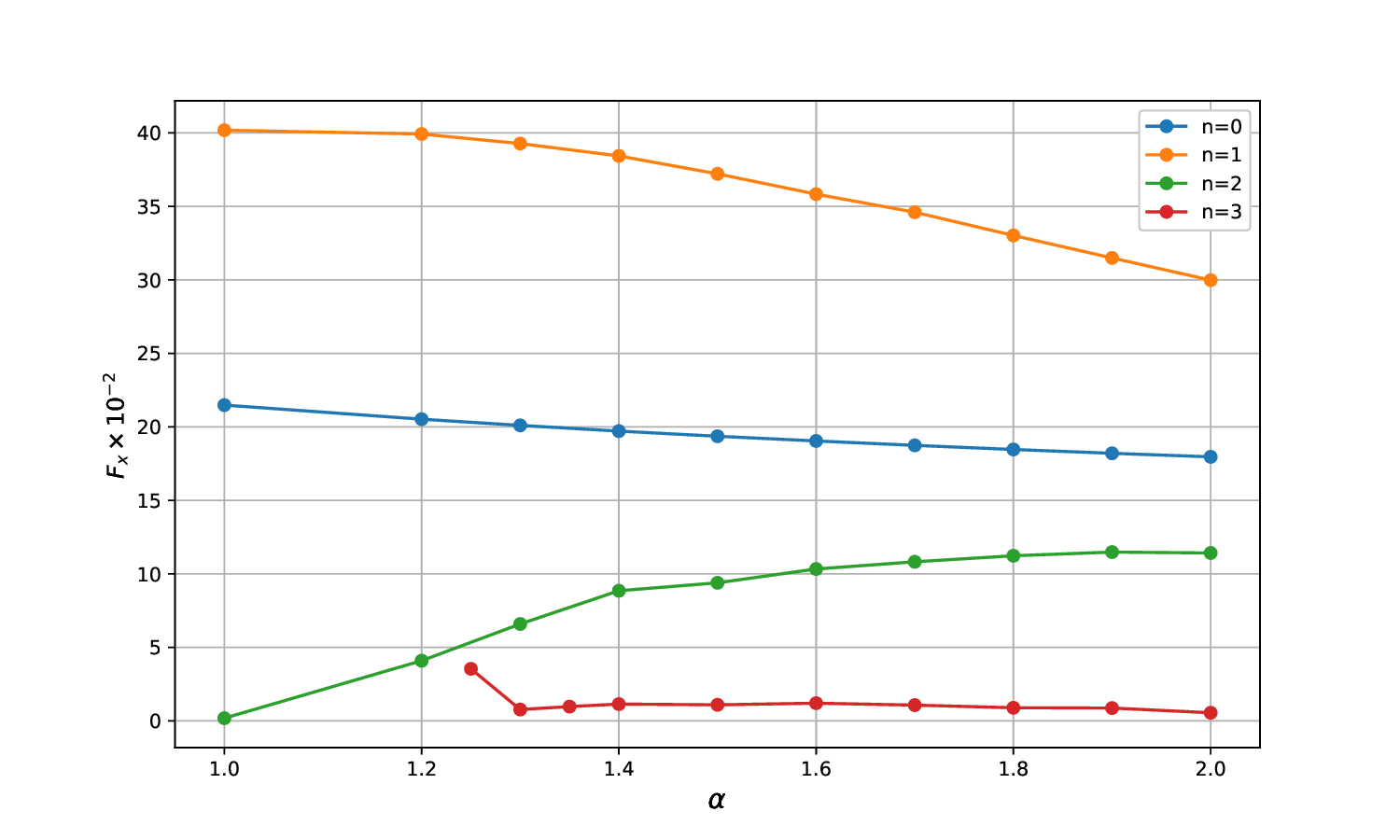}
\par\end{centering}
\caption{Fisher information $F_x$ versus $\alpha$ for four levels $n=0,1,2,3$ }
\label{fig6}
\end{figure}

\begin{figure}
\begin{centering}
\includegraphics[scale=0.5]{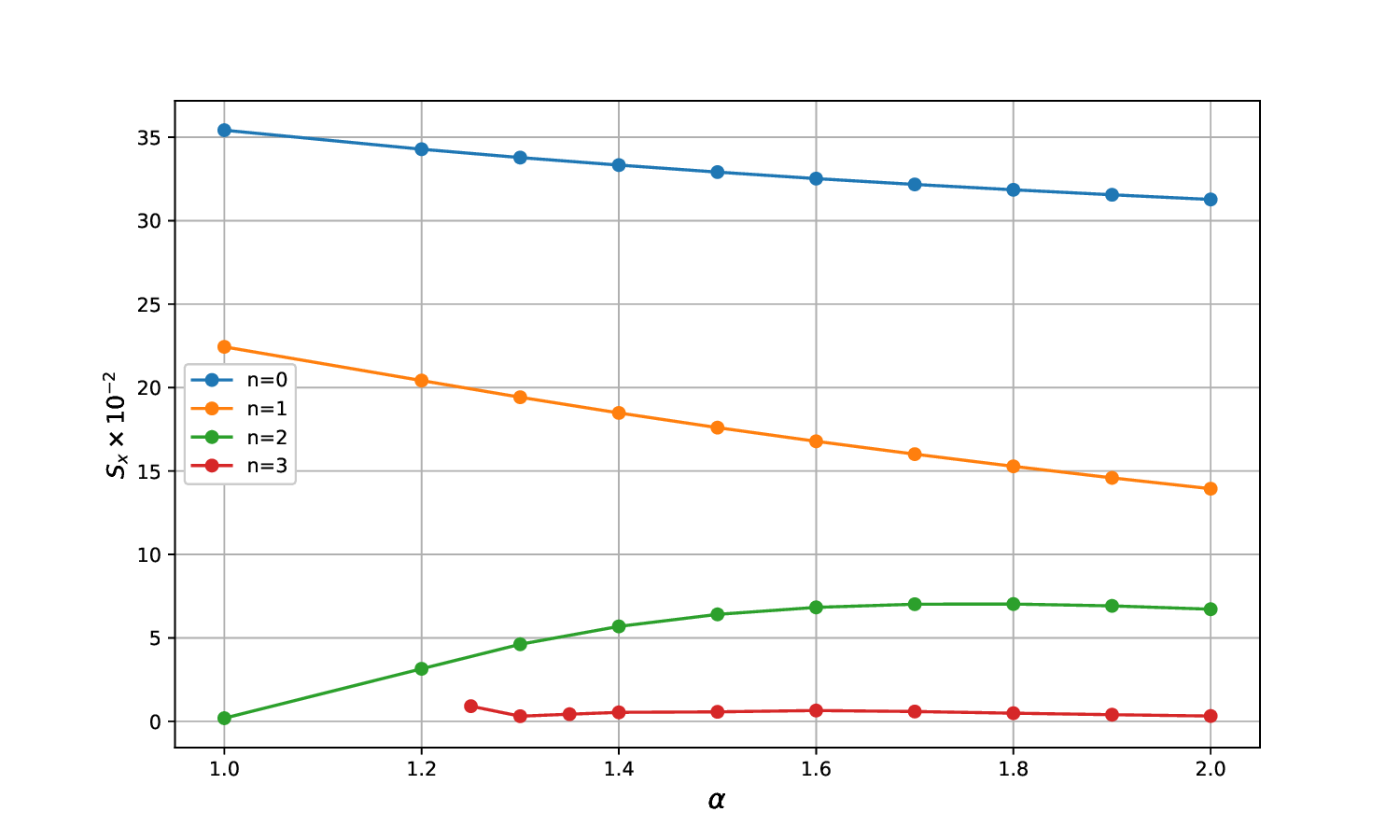}
\par\end{centering}
\caption{Shannon entropy $S_x$ versus $\alpha$ for four levels $n=0,1,2,3$ }
\label{fig7}
\end{figure}

Fisher information \( (F) \) and Shannon entropy \( (S) \) in coordinate space for the specific state of the confined harmonic oscillator are defined in terms of the probability density \( \rho_n(x) = \left|\psi_n(x)\right|^2 \) as follows:
\begin{equation}
F_n(x) = \int \frac{\left|\nabla \rho_n(x)\right|^2}{\rho_n(x)} dx, \label{eq:b15}
\end{equation}
and
\begin{equation}
S_n(x)=-\int\rho_n(x)\ln\rho_n(x)dx.\label{eq:12}
\end{equation}
Their extension in the fractional derivative framework is
\begin{equation}
F_{n,\alpha}(x) = \int \frac{\left|\nabla \rho_{n,\alpha}(x)\right|^2}{\rho_{n,\alpha}} dx, \label{eq:b15}
\end{equation}
\begin{equation}
S_{n,\alpha}(x)=-\int\rho_\alpha(x)\ln\rho_\alpha(x)dx.\label{eq:13}
\end{equation}
The explicit derivation of these quantities is generally challenging. The calculation of these parameters encounters significant obstacles due to the logarithmic factors in the integrals, rendering analytical expressions nearly impossible to obtain. To address these challenges, we employ two strategies: (i) representation of Shannon and Fisher information entropy densities, and (ii) numerical computation of the integrals. The first step ensures that both parameters remain positive. Based on the probability density figures, we observe that \(\ln\rho_n(x) \leq 0\). Consequently, Figures \ref{fig4} and \ref{fig5} demonstrate that the density of Fisher and Shannon information for both \( n = 0 \) (left) and \( n = 1 \) (right) across different values of \( \alpha \) are always positive. This observation allows us to access the Fisher and Shannon parameters reliably.
The results of the calculations of these parameters are presented in Figures \ref{fig6} and \ref{fig7}. It is important to note that in these figures, both \(n\) and \(\alpha\) have been varied.

The Fisher information, which quantifies the amount of information that an observable random variable conveys about an unknown parameter, can be significantly influenced by the fractional nature of a system's dynamics. Fractional derivatives, which encapsulate the memory and hereditary characteristics of processes, affect the gradient and curvature of probability density functions, thereby altering the Fisher information. Similarly, Shannon entropy, a measure of uncertainty or randomness in a system, is also affected by fractional dynamics. The nonlocal properties of fractional derivatives influence the distribution and spread of probability densities, leading to changes in the calculated entropy. This is particularly pertinent in systems where probability densities exhibit heavy tails or long-range correlations, phenomena naturally described by fractional calculus.

In this context, having determined these parameters for varying values of \(\alpha\), we are prepared to compute specific quantities within the framework of statistical measures in quantum systems. Several indicators, developed within the realms of information theory and complexity theory—such as Fisher information, Shannon entropy, and statistical measures of complexity—have been calculated for various systems using different approaches (see Ref. and the references therein for further details). The probability densities that characterize the state of a quantum system are defined in both position and momentum spaces. From these densities, the statistical complexity and Fisher–Shannon information are derived.

At this juncture, we introduce the López-Ruiz, Mancini, and Calbet (LMC) complexity measure, denoted as \( C \) and defined by the equation:
\[
C = H \cdot D
\]
where \( D \) represents the disequilibrium of the system, measuring the concentration of the spatial distribution:
\[
D = \int \rho^2(x) \, dx,
\]
and \( H \) is closely related to the Shannon entropy, specifically defined as the exponential Shannon entropy:
\[
H = \mathrm{e}^{S}.
\]
The concept of complexity, initially proposed by López-Ruiz et al., posits that complexity \( C \) reaches its peak when a system is in an intermediate state between perfect order and complete disorder, characterized by significant values of both information \( H \) and disequilibrium \( D \). This approach effectively distinguishes between simple and complex systems—such as a perfect crystal and an ideal gas—by integrating the system's informational content with the probabilistic distribution of its states. Consequently, complexity \( C \) emerges as a composite measure that combines entropy \( H \) with disequilibrium \( D \), thereby quantifying the organization of information within a system and providing insights into its structural organization and stability.

Disequilibrium, in particular, is defined as the deviation of a system's state distribution from equiprobability. In a perfect crystal, where a single state dominates, disequilibrium is high. Conversely, in an ideal gas, where all states are equally probable, disequilibrium is minimal. The interplay between information \( H \) and disequilibrium \( D \) forms the basis of the complexity measure, capturing the hierarchical structure of probabilities within the system.

Another parameter introduced alongside complexity \( C \) is the Fisher-Shannon information measure \( P \). This parameter offers further insights into the complexity of a system. The measure \( P \) complements the evaluation of complexity by incorporating aspects of both Fisher information and Shannon entropy, thereby enriching the overall understanding of the system's informational structure. The Fisher-Shannon information \( P \) is formulated as:
\[
P = J \cdot F,
\]
where \( J \) is related to the Shannon entropy and is defined by:
\[
J = \frac{1}{2\pi e} \mathrm{e}^{2S/3},
\]
and \( F \) denotes the Fisher information. The product \( P \) thus integrates the spread of the probability distribution (represented by \( J \)) with the precision or sharpness of the distribution (represented by \( F \)). This dual consideration provides a comprehensive framework for analyzing complexity and information measures within quantum systems. It underscores the utility of various statistical indicators in assessing the hierarchical structure and underlying dynamics of these systems, ultimately advancing the understanding of quantum mechanics and its practical applications.

\begin{figure}
\begin{centering}
\includegraphics[scale=0.5]{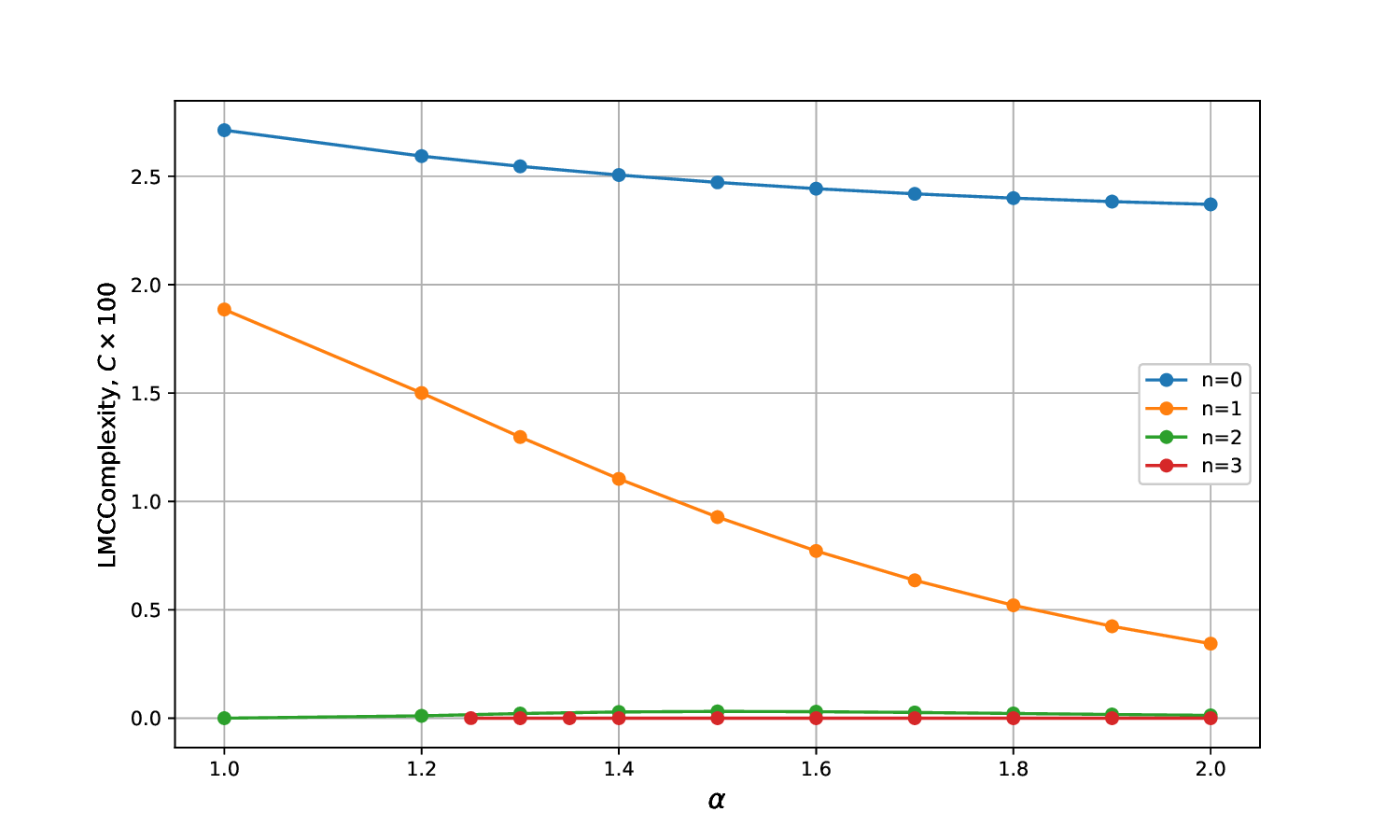}
\par\end{centering}
\caption{ LMC complexity $C$  versus $\alpha$ for four levels $n=0,1,2,3$ }
\label{fig8}
\end{figure}

\begin{figure}
\begin{centering}
\includegraphics[scale=0.5]{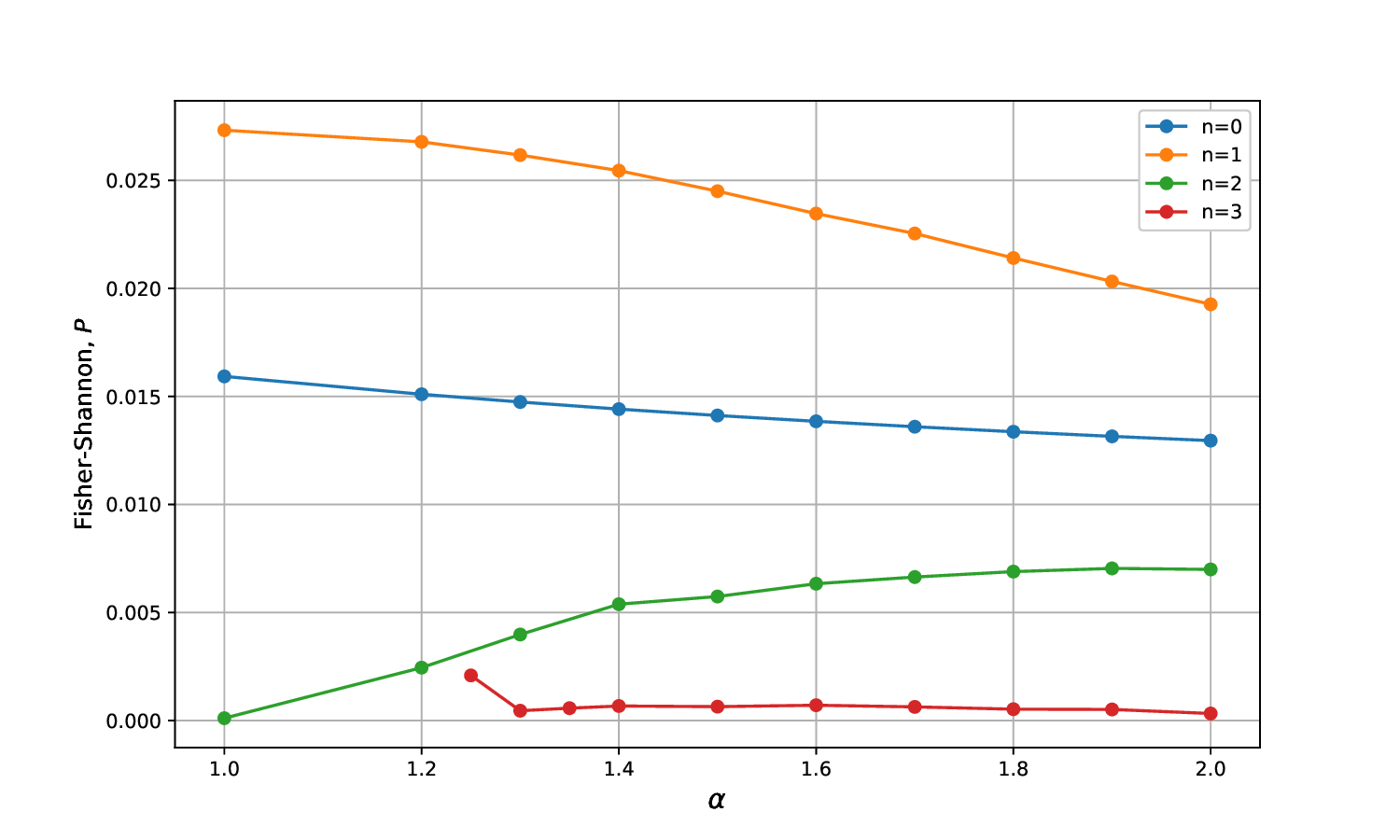}
\par\end{centering}
\caption{ Fisher-Shannon Information $P$ versus $\alpha$ for four levels $n=0,1,2,3$ }
\label{fig9}
\end{figure}
Figures \ref{fig8} and \ref{fig9} depict the parameters \( C \) (complexity) and \( P \) (Fisher-Shannon information), both of which provide important insights into the physical properties of the system. The complexity parameter \( C \) reflects the degree of disorder or structural intricacy within the system, while the Fisher-Shannon information \( P \) quantifies the system's information content and its sensitivity to perturbations.

As shown in Figure \ref{fig8}, the complexity parameter \( C \) for levels \( n = 0, 1 \) reveals an increase in disorder as the system transitions from the relativistic regime (\(\alpha = 1\)) to the non-relativistic regime (\(\alpha = 2\)). This suggests that at lower energy levels, the system exhibits greater structural complexity in the relativistic regime. In contrast, at higher levels \( n = 2, 3 \), the complexity \( C \) approaches zero, indicating a lack of disorder and a more stable configuration in these cases (refer to Table I for additional details).

Figure \ref{fig9} presents the Fisher-Shannon information \( P \) as a function of the parameter \(\alpha\) for four levels \( n = 0, 1, 2, 3 \). The decreasing trend of \( P \) with increasing levels suggests that as the energy level rises, the system becomes less sensitive to fluctuations, implying a reduction in informational content and structural complexity.
\begin{table}[H]
\caption{LMC Complexity $C$ and Fisher-Shannon Information $P$ for four levels $n=0,1,2,3$}
\begin{centering}
\begin{tabular}{|c|c|c|c|}
\hline 
$n=0$ & $n=1$ & $n=2$ & $n=3$\tabularnewline
\hline 
\hline 
\begin{tabular}{|c|c|c|}
\hline 
$\alpha$ & $C$ & $P$\tabularnewline
\hline 
1.0 & 2.7126 & 0.0159 \tabularnewline
1.2 & 2.5931 & 0.0151 \tabularnewline
1.3 & 2.5459 & 0.0147 \tabularnewline
1.4 & 2.5059 & 0.0144 \tabularnewline
1.5 & 2.4718 & 0.0141 \tabularnewline
1.6 & 2.4429 & 0.0138 \tabularnewline
1.7 & 2.4190 & 0.0136 \tabularnewline
1.8 & 2.3993 & 0.0134 \tabularnewline
1.9 & 2.3832 & 0.0132 \tabularnewline
2.0 & 2.3704 & 0.0129 \tabularnewline
\hline 
\end{tabular} & 
\begin{tabular}{|c|c|c|}
\hline 
$\alpha$ & $C$ & $P$\tabularnewline
\hline 
1.0 & 1.8855 & 0.0273 \tabularnewline
1.2 & 1.5004 & 0.0268 \tabularnewline
1.3 & 1.2972 & 0.0262 \tabularnewline
1.4 & 1.1041 & 0.0255 \tabularnewline
1.5 & 0.9278 & 0.0245 \tabularnewline
1.6 & 0.7715 & 0.0235 \tabularnewline
1.7 & 0.6361 & 0.0225 \tabularnewline
1.8 & 0.5208 & 0.0214 \tabularnewline
1.9 & 0.4241 & 0.0203 \tabularnewline
2.0 & 0.3439 & 0.0193 \tabularnewline
\hline 
\end{tabular} & 
\begin{tabular}{|c|c|c|}
\hline 
$\alpha$ & $C$ & $P$\tabularnewline
\hline 
1.0 & 2.8444e-05 & 0.0001 \tabularnewline
1.2 & 0.0108 & 0.0024 \tabularnewline
1.3 & 0.0215 & 0.0040 \tabularnewline
1.4 & 0.0288 & 0.0054 \tabularnewline
1.5 & 0.0313 & 0.0057 \tabularnewline
1.6 & 0.0300 & 0.0063 \tabularnewline
1.7 & 0.0264 & 0.0066 \tabularnewline
1.8 & 0.0218 & 0.0069 \tabularnewline
1.9 & 0.0171 & 0.0070 \tabularnewline
2.0 & 0.0129 & 0.0070 \tabularnewline
\hline 
\end{tabular} & 
\begin{tabular}{|c|c|c|}
\hline 
$\alpha$ & $C$ & $P$\tabularnewline
\hline 
1.25 & 2.0319e-04 & 0.0021 \tabularnewline
1.3 & 9.3262e-06 & 0.0005 \tabularnewline
1.35 & 1.4391e-05 & 0.0006 \tabularnewline
1.4 & 1.9213e-05 & 0.0007 \tabularnewline
1.5 & 1.5892e-05 & 0.0006 \tabularnewline
1.6 & 1.7154e-05 & 0.0007 \tabularnewline
1.7 & 1.1981e-05 & 0.0006 \tabularnewline
1.8 & 7.0477e-06 & 0.0005 \tabularnewline
1.9 & 3.8307e-06 & 0.0005 \tabularnewline
2.0 & 1.9994e-06 & 0.0003 \tabularnewline
\hline 
\end{tabular}\tabularnewline
\hline 
\end{tabular}
\par\end{centering}
\end{table}

\section{Conclusion}

This study has successfully applied the Riesz-Feller fractional derivative to analyze the Fisher and Shannon information measures for a one-dimensional fractional quantum harmonic oscillator. By extending traditional quantum mechanics into the fractional domain, we have unveiled significant insights into the probabilistic nature of quantum systems under fractional calculus. The results illustrate how the fractional parameter \( \alpha \) influences the Fisher information and Shannon entropy, which are critical for understanding the quantum system's informational characteristics.

Our findings indicate that the fractional dynamics considerably affect both the Fisher information, which measures the precision of parameter estimation in quantum systems, and the Shannon entropy, which quantifies the uncertainty inherent in these systems. The study further extends the analysis by incorporating statistical measures of complexity, such as the López-Ruiz, Mancini, and Calbet (LMC) complexity, and Fisher-Shannon information, providing a comprehensive framework for evaluating the complexity and information content of fractional quantum systems.

Overall, this research contributes to the broader field of quantum information theory by highlighting the utility of fractional calculus in exploring quantum mechanics' foundational aspects. The methodologies and results presented here could serve as a basis for further investigations into the application of fractional calculus in various quantum systems, potentially leading to new theoretical advancements and practical applications in quantum mechanics.

\appendix

\section{The Fractional Factorization Method Based on Riesz-Feller Derivatives}

We begin with the following equation expressed in the momentum representation:

\begin{equation}
\left[\left|k\right|^{\alpha} - \frac{d^{2}}{dk^{2}}\right]\phi(k) = E\phi(k), \label{eq:b3}
\end{equation}

where \(\phi(k)\) represents the Fourier transform of \(\psi(x)\). The factorization algorithm for the fractional quantum harmonic oscillator is subsequently applied as proposed by Olivar-Romero and Rosas-Ortiz \cite{OlivarRomeroF2016}. Consider a pair of operators \( A_{\alpha} \) and \( B_{\alpha} \) such that:

\begin{equation}
H_{\alpha} = B_{\alpha}A_{\alpha} + \epsilon_{\alpha}, \label{eq:b4}
\end{equation}

where \(\epsilon_{\alpha}\) may represent a fractional-differential operator. The operators \( A_{\alpha} \), \( B_{\alpha} \), and \(\epsilon_{\alpha}\) are defined as:

\begin{equation}
A_{\alpha} = \frac{d^{\alpha/2}}{dx^{\alpha/2}} + x, \label{eq:b5}
\end{equation}
\begin{equation}
B_{\alpha} = -\frac{d^{\alpha/2}}{dx^{\alpha/2}} + x, \label{eq:b6}
\end{equation}
\begin{equation}
\epsilon_{\alpha} = \frac{\alpha}{2} \frac{d^{\alpha/2-1}}{dx^{\alpha/2-1}}. \label{eq:b7}
\end{equation}

Equation \eqref{eq:b7} demonstrates that for \(\alpha \neq 2\), the factorization remainder \(\epsilon_{\alpha}\) is a fractional differential operator of order \(\alpha/2 - 1\). Conversely, when \(\alpha = 2\), it simplifies to the constant \(\epsilon_{2} = 1/2\), with the operators \( A_{2} \) and \( B_{2} \) corresponding to the conventional annihilation and creation operators of the standard harmonic oscillator.

To find the kernel of \( A_{\alpha} \), we solve the following fractional differential equation:

\begin{equation}
A_{\alpha}\psi_{1,0}^{\alpha} = \left(-\frac{d^{\alpha/2}}{dx^{\alpha/2}} + x\right)\psi_{0}^{\alpha} = 0. \label{eq:b8}
\end{equation}

In the momentum representation, Equation \eqref{eq:b8} transforms into:

\begin{equation}
i \left[\text{sign}(k)k^{\alpha/2} + \frac{d}{dk}\right]\phi_{0}^{\alpha} = 0. \label{eq:b9}
\end{equation}

The solution to Equation \eqref{eq:b9} is:

\begin{equation}
\phi_{0}^{\alpha} = e^{-\frac{2|k|^{\alpha/2+1}}{\alpha+2}}. \label{eq:b10}
\end{equation}

Analogous to traditional factorization, applying the operators \( A_{\alpha} \) and \( B_{\alpha} \) appropriately yields additional solutions. The solutions for the first few excited states are as follows \cite{OlivarRomeroF2016}:

\begin{itemize}
\item For the first excited state \(\phi_{1}^{\alpha}(k)\):
\begin{equation}
\phi_{1}^{\alpha}(k) = -2i |k|^{\alpha} \text{sgn}(k) e^{-\frac{2|k|^{\alpha/2+1}}{\alpha+2}}. \label{eq:b11}
\end{equation}

\item For the second excited state \(\phi_{2}^{\alpha}(k)\):
\begin{equation}
\phi_{2}^{\alpha}(k) = \left(\alpha |k|^{\alpha/2-1} - 4 |k|^{\alpha}\right)e^{-\frac{2|k|^{\alpha/2+1}}{\alpha+2}}. \label{eq:b12}
\end{equation}

\item For the third excited state \(\phi_{3}^{\alpha}(k)\) \cite{Rosu2020}:
\begin{equation}
\phi_{3}^{\alpha}(k) = i \text{sgn}(k) \left(-8 |k|^{3\alpha/2} + 6 \alpha |k|^{\alpha-1} - \alpha \left(\frac{\alpha}{2} - 1\right) |k|^{\alpha/2-2}\right)e^{-\frac{2|k|^{\alpha/2+1}}{\alpha+2}}. \label{eq:b13}
\end{equation}
\end{itemize}

In general, one can express the solution as:

\begin{equation}\label{e-phigen}
\phi_n(k) = i^n\widetilde{H}_n\phi_{0}^{(\alpha)},
\end{equation}

where \(\widetilde{H}_n(k)\) are the fractionally-deformed Hermite 'polynomials':

\begin{eqnarray}
&\widetilde{H}_0=1, \nonumber\\
&\widetilde{H}_1(k)=2~\mathrm{sgn}(k)|k|^{\frac{\alpha}{2}}, \nonumber\\
&\widetilde{H}_2(k)=4|k|^{\frac{2\alpha}{2}}-\alpha|k|^{\frac{\alpha}{2}-1}, \nonumber\\
&\widetilde{H}_3(k)=\mathrm{sgn}(k)\bigg[8| k|^{\frac{3\alpha}{2}}-6\alpha|k|^{\frac{2\alpha}{2}-1}+2\frac{\alpha}{2}\left(\frac{\alpha}{2}-1\right)|k|^{\frac{\alpha}{2}-2}\bigg], \nonumber\\
&\widetilde{H}_4(k)=16|k|^{\frac{4\alpha}{2}}-24\alpha |k|^{\frac{3\alpha}{2}-1}+6\alpha(\alpha-1)|k|^{\frac{2\alpha}{2}-2} \nonumber\\
&+2\frac{\alpha}{2}\left(\frac{\alpha}{2}-1\right)|k|^{\frac{2\alpha}{2}-2}-2\frac{\alpha}{2}\left(\frac{\alpha}{2}-1\right)\left(\frac{\alpha}{2}-2\right)|k|^{\frac{\alpha}{2}-3}, \nonumber\\
&\vdots
\end{eqnarray}

which we refer to as Riesz-Feller Hermite 'polynomials.' For \(\alpha = 2\), these reduce to the standard Hermite polynomials, with a sign change for the odd polynomials, albeit in the \(|k|\) variable. 
The general form for \(\widetilde{H}_n(k)\) is:

\begin{eqnarray}\label{genH}
\widetilde{H}_n(k) &= \mathrm{sgn}(k)^n \Big[ 2^n|k|^{\frac{n\alpha}{2}} - p_1(\alpha)|k|^{\frac{(n-1)\alpha}{2}-1} + p_2(\alpha)|k|^{\frac{(n-2)\alpha}{2}-2} \nonumber\\
&- p_3(\alpha)|k|^{\frac{(n-3)\alpha}{2}-3} + \dots + (-1)^{n-1}p_{n-1}(\alpha)|k|^{\frac{\alpha}{2}-(n-1)}\Big],
\end{eqnarray}

where \(p_i(\alpha)\) are polynomials of order \(i\) in \(\alpha\), which can be derived from the following counterpart of the Rodrigues formula:

\begin{equation}\label{rodr}
\widetilde{H}_{n}(k) =
(-1)^n{\rm sgn}(k)^n\,e^{2\frac{|k|^{\frac{\alpha}{2}+1}}{\frac{\alpha}{2}+1}}\frac{d^n}{dk^n}e^{-2\frac{|k|^{\frac{\alpha}{2}+ 1}}{\frac{\alpha}{2}+1}},
\end{equation}

which, for \(\alpha = 2\), simplifies to:

\begin{equation}\label{rodr2}
\widetilde{H}_{n}(k) = (-1)^n{\rm sgn}(k)^n\, e^{|k|^{2}}\frac{d^n}{dk^n}e^{-|k|^{2}},
\end{equation}

to be compared with the standard \(x\) space formula:

\[
H_{n}(x) = (-1)^n\, e^{x^{2}}\frac{d^n}{dx^n}e^{-x^{2}}.
\]
```

\bibliography{korichi}% Produces the bibliography via BibTeX.
\end{document}